\pdfoutput=1
\documentclass[twocolumn, tighten]{aastex62}
\usepackage{amsmath}

\shorttitle{The year-long flux variation in Boyajian's star are asymmetric or aperiodic}
\shortauthors{Hippke \& Angerhausen}

\begin{document}
\title{The year-long flux variations in Boyajian's star are asymmetric or aperiodic}

\author[0000-0002-0794-6339]{Michael Hippke}
\affiliation{Sonneberg Observatory, Sternwartestr. 32, 96515 Sonneberg, Germany}
\email{michael@hippke.org}

\author[0000-0001-6138-8633]{Daniel Angerhausen}
\affiliation{Center for Space and Habitability, University of Bern, Hochschulstrasse 6, 3012 Bern, Switzerland}
\affiliation{Blue Marble Space Institute of Science,
1001 4th ave, Suite 3201
Seattle, Washington 98154
USA}
\email{daniel.angerhausen@csh.unibe.ch}

\begin{abstract}
We combine and calibrate publicly available data for Boyajian's star including photometry from ASAS (SN, V, I), Kepler, Gaia, SuperWASP, and citizen scientist observations (AAVSO, HAO and Burke-Gaffney). Precise (mmag) photometry covers the years $2006-2017$. We show that the year-long flux variations with an amplitude of $\approx4$\,\% can not be explained with cyclical symmetric or asymmetric models with periods shorter than ten years. If the dips are transits, their period must exceed ten years, or their structure must evolve significantly during each 4-year long cycle.\\
\end{abstract}

\section{Introduction}
Boyajian’s Star (KIC~8462852) is a mysterious object which showed asymmetric, aperiodic day-long deep (20\,\%) dips in brightness during Kepler's $2009-2013$ mission \citep{2016MNRAS.457.3988B}. The mystery deepened when \citet{2016ApJ...822L..34S} claimed a dimming of the star during $1890-1990$ based on historical plates, and \citet{2016ApJ...830L..39M} showed that the dimming continued during Kepler's mission. These results have been interpreted that the brightness of Boyajian’s Star is monotonically decreasing with time. Although the century-long dimming has been challenged in re-analyses \citep{2016ApJ...825...73H,2016arXiv160502760L} and with data from multiple other observatories \citep{2017ApJ...837...85H}, it only recently became clear that the star's brightness shows reoccurring variations with a few percent amplitude on 8-year long timescales \citep{2017arXiv170807822S}.

Dimmings and variations are common for young stars \citep{2017MNRAS.470..202B}, but are not known for F3 main sequence stars \citep{2016MNRAS.457.3988B}. A series of more or less exotic solutions have been proposed, such as the ingestion of a planet \citep{2017MNRAS.468.4399M}, intrinsic variations \citep{2017ApJ...842L...3F} or Solar System debris \citep{2017MNRAS.471.3680K}. To narrow down the model choice, we here examine available data to determine the brightness variations over the last decade.

\section{Method}

\begin{figure}
\includegraphics[width=\linewidth]{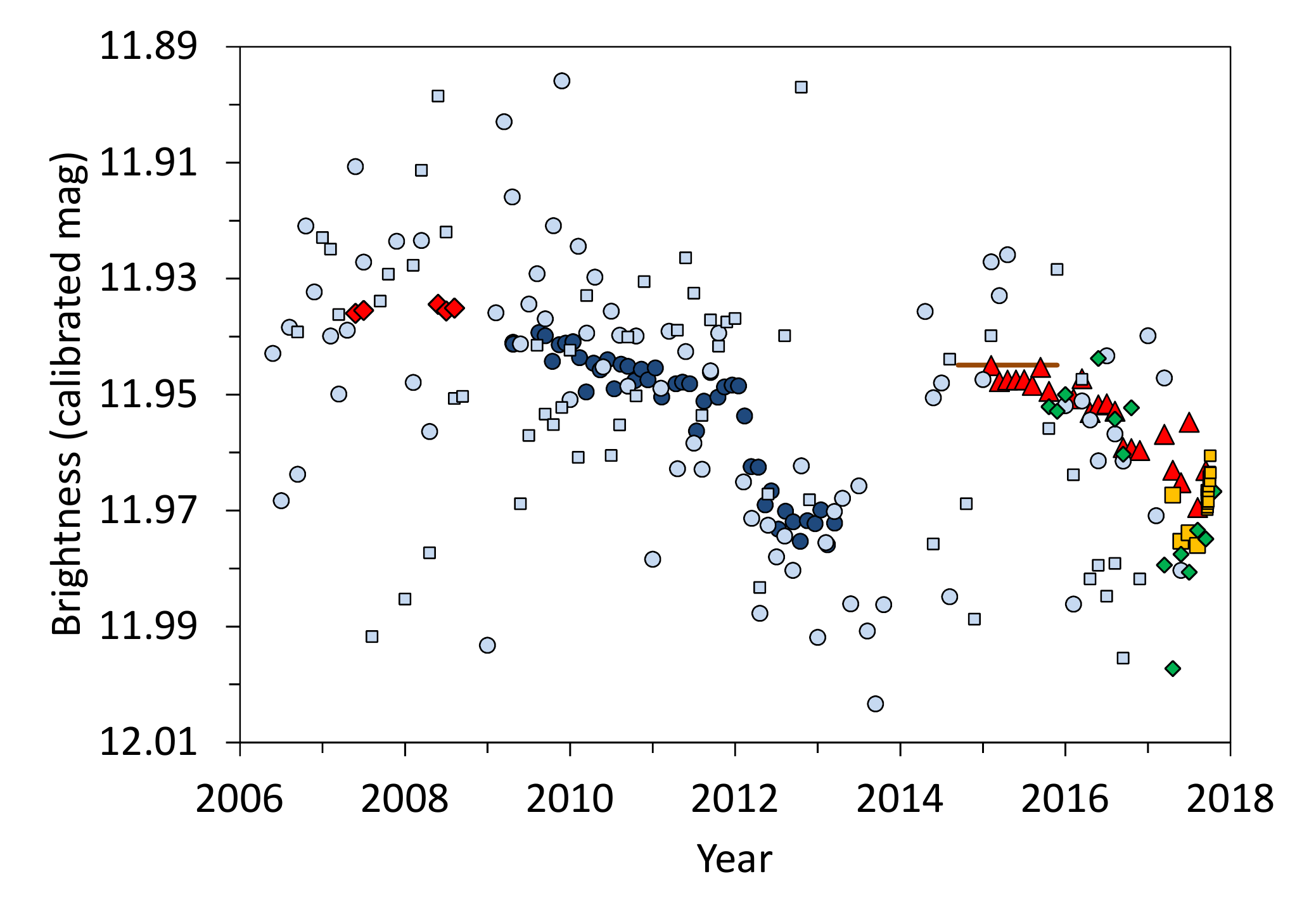}
\caption{\label{fig1}Timeseries overview of all data used in this analysis: ASAS-I (light blue squares), ASAS-SN (red triangles), SuperWASP (red diamonds), Gaia (line), AAVSO and LDJ (green diamonds) and HAO (orange squares).}
\end{figure}

\subsection{Observations}
Several independent datasets cover the brightness of Boyajian's star over different intervals. Before the release of the ASAS data, the creation of a consistent light curve was not possible, because few of the other datasets overlap and offsets due to calibration differences were uncertain.

\subsubsection{ASAS-V and ASAS-I}
ASAS is a long-term V- and I-band wide-field variability survey based in Chile (since 1997) and on Hawaii (since 2006) \citep{1997AcA....47..467P,2002AcA....52..397P}. For Boyajian's Star, 486 good quality observations were obtained in V- and 247 in I-band between May 2006 and May 2017. Compared to ASAS-SN, ASAS has a longer time baseline, but larger photometric uncertainty (0.028\,mag per observation). This is due to the smaller apertures (5\,cm versus 14\,cm) and lower cadence.

\subsubsection{ASAS-SN}
ASAS-SN is an all-sky V-band transient survey at the same location as ASAS \citep{2014ApJ...788...48S}. For Boyajian's Star, 377 observations were taken between February 2015 and May 2017. Aperture photometry was performed by \citet{2017arXiv170807822S} with a median photometric uncertainty of 0.01\,mag per observation.

\subsubsection{SuperWASP}
The WASP was a wide-field white-light survey located on the Canary Islands and South Africa \citep{2006PASP..118.1407P}. For Boyajian's Star, a total of 5,377 measurements were taken over three observing seasons ($2006-2008$). The first season with only 22 observations shows  0.2\,mag offsets for many stars, and we discard these data. The remaining data cover the time intervals $2007.41-2007.51$ and $2008.36-2008.59$.

\subsubsection{Kepler FFI}
Kepler photometry is optimized to detect small, short duration signals such as planet transits at the expense of long-term trends. Long-term variability ($>30$ days) can be recovered in photometry from the Full Frame Images (FFIs), as shown by \citet{2017arXiv170507928M,2016ApJ...830L..39M,2017ascl.soft05006M}. A total of 52 such measurements could be extracted for Boyajian's Star, with a typical uncertainty of 0.001\,mag.

\subsubsection{Citizen science observations}
Citizen scientists from the American Association of Variable Star Observers (AAVSO) collected $>30,000$ V-band observations between 2015.7 and 2017.8 with varying quality. We tried different quality cuts such as only taking observations with low estimated uncertainties or only observations from observers with a minimum number of images. Compared to the full dataset, all tested quality cuts yield very similar results within a few mmag when taking monthly bins. We choose to only eliminate observations that differ by more than $5\sigma$ from the mean of a month, and show the monthly bins in Figure~\ref{fig1}. 

Citizen scientist Bruce Gary of Hereford Arizona Observatory publicly released 75 observations\footnote{\url{http://www.brucegary.net/ts4/}} mostly in V-band taken between 2017.3 and 2017.8 with a typical uncertainty of 0.002\,mag. Starting in October 2017, observations were taken in g'-band with a date-average uncertainty of 0.001 mag.

Citizen scientist Dave Lane (LDJ) of Burke-Gaffney Observatory provided 520 observations taken with a Planewave CDK24 telescope (aperture 0.61\,m, focal length 3970\,mm) located in Halifax, Canada with an Apogee CG16M CCD camera (KAF-16803 sensor binned 2x2) and a 50\,mm square Astrodon V filter. Observations from May 1, 2016 to September 23, 2017 (HJD  2457509.6 to 2458019.6) were differential photometry using the single AAVSO comparison star 000-BLS-549 (mag=12.427). Subsequent observations are ensemble photometry using five AAVSO comparison stars.

\subsubsection{Gaia}
Variability information is not released in Gaia's DR1 \citep{2016A&A...595A...2G}, but it has been shown by \citet{2017MNRAS.466.4711B} that variability can be approximated from the average measured flux and the reported uncertainty. As shown by \citet{2017arXiv170807822S}, Boyajian's star is less variable than 11 of 14 comparison stars between July 2014 and September 2015, which have an average scatter of 1.6\,\%. We therefore show the Gaia data as a line in Figure~\ref{fig1}, noting that the width of this line is uncertain until Gaia's DR4 which will include the photometry.

\subsection{Normalization}
In order to compare all datasets to each other and eventually combine them to one overall timeseries a benchmark is needed.  Fortunately the ASAS-V and ASAS-I cover the entire timeline from 2006 to the present and can be used a reference zero point value for the other datasets. A comparison of V to I shows that the I-band has fewer observations, larger uncertainties and is less affected from the brightness variations we seek to analyze here. Therefore, we choose the ASAS-V data as our baseline. The other datasets are calibrated by normalizing their flux values with respect to ASAS-V, so that the squared residuals are minimized for the overlapping segments. We tested the nearest-neighbor method (in time), splines, and monthly bins and got virtually identical results. We proceed by using the calibration to monthly bins as shown in Figure~\ref{fig1} for maximum simplicity. The combined light curve suffers from bandpass differences. We assume, however, that color-differences during the variations have amplitudes that are on much smaller scales than the overall brightness changes. For example \citet{2018arXiv180100732B} and \citet{2018arXiv180100720D} showed that even for the strong short term dimmings, that make Kic 846 so peculiar, the difference between very different bandpasses such as r and B is only a few tenth of a percent while all bandpasses clearly show the dimmings on several percent level (Fig. 7 in their paper). Most of our data consists of V or white light measurements that we use to analyze long term trends and cyclical variation. In general, as our results rely on the qualitative fit, slight differences at the mmag level are irrelevant. While calculations were made on the individual data points, we show monthly bins in the Figures for better visibility. Despite the larger scatter in ASAS-V and ASAS-I compared to the other datasets, their baseline is helpful because they span the entire time from 2006-2017.

 Using ASAS data, \citet{2017arXiv170807822S} found indications for cyclical brightness variations with a period of $\approx8$ years. The variability is color-dependent. The dimming is less in UV ($0.2\,\mu$m) than IR ($4.5\,\mu$m), so that the responsible bodies must be small (microns) in size \citep{2017ApJ...847..131M}. The brightening seen in ASAS-V is not clearly visible in ASAS-I, also noted by \citet{2017arXiv170807822S}. These bandpass variations are also evident in our combined light curve, so that the true amplitudes may be incorrect by up to 20\,\%, depending on color.

There is also a significant overlap between ASAS-SN, AAVSO, and HAO; and the gap between SuperWASP and Kepler is short (8 months). As can be seen in Figure~\ref{fig1}, all datasets can consistently be co-added into a combined light curve. The systematic calibration uncertainty is small: The SuperWASP average is 11.935\,mag when calibrated to ASAS-V and would be 11.941\,mag when taking the closest Kepler point as reference, a difference of 0.6\,\% in brightness. A similar uncertainty exists for the ASAS-SN, AAVSO, and HAO datasets at the level of 0.1--0.3\,\%. These variations are mostly due to the bandpass differences and do not affect our qualitative results.

\section{Results and Model Discussion}
In Figure \ref{fig1} we show the 2006-2017 longterm photometry of KIC~8462852. Using the ASAS-V monitoring data as baseline we were able to combine all other photometric data available to us at the time of submission. The plot shows various levels of variation which we discuss deeper in the following subsections.

\begin{figure}
\includegraphics[width=\linewidth]{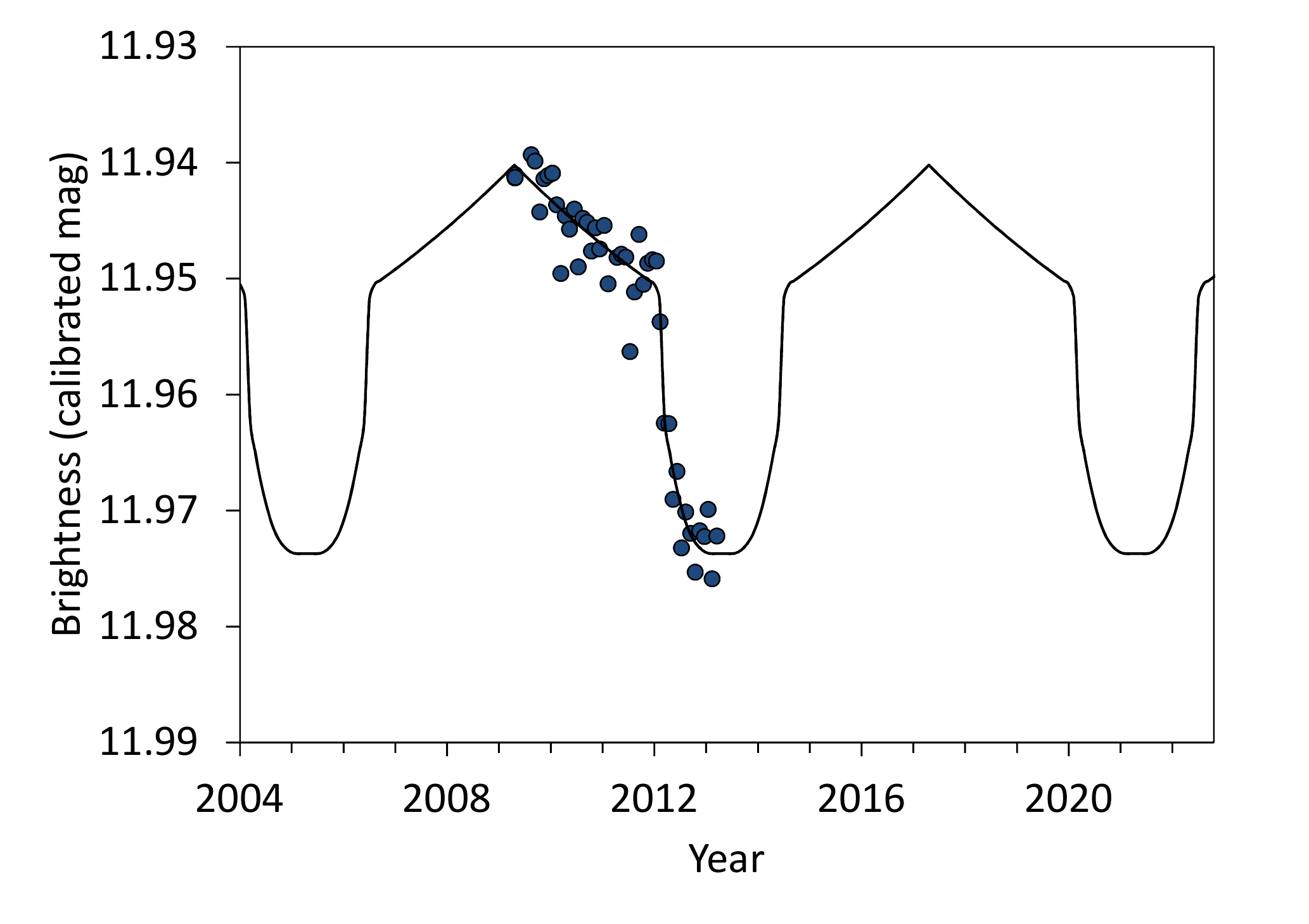}
\caption{\label{fig2}Shortest possible period symmetrical model (line) using Kepler data (blue circles).}
\end{figure}

\begin{figure*}
\includegraphics[width=.5\linewidth]{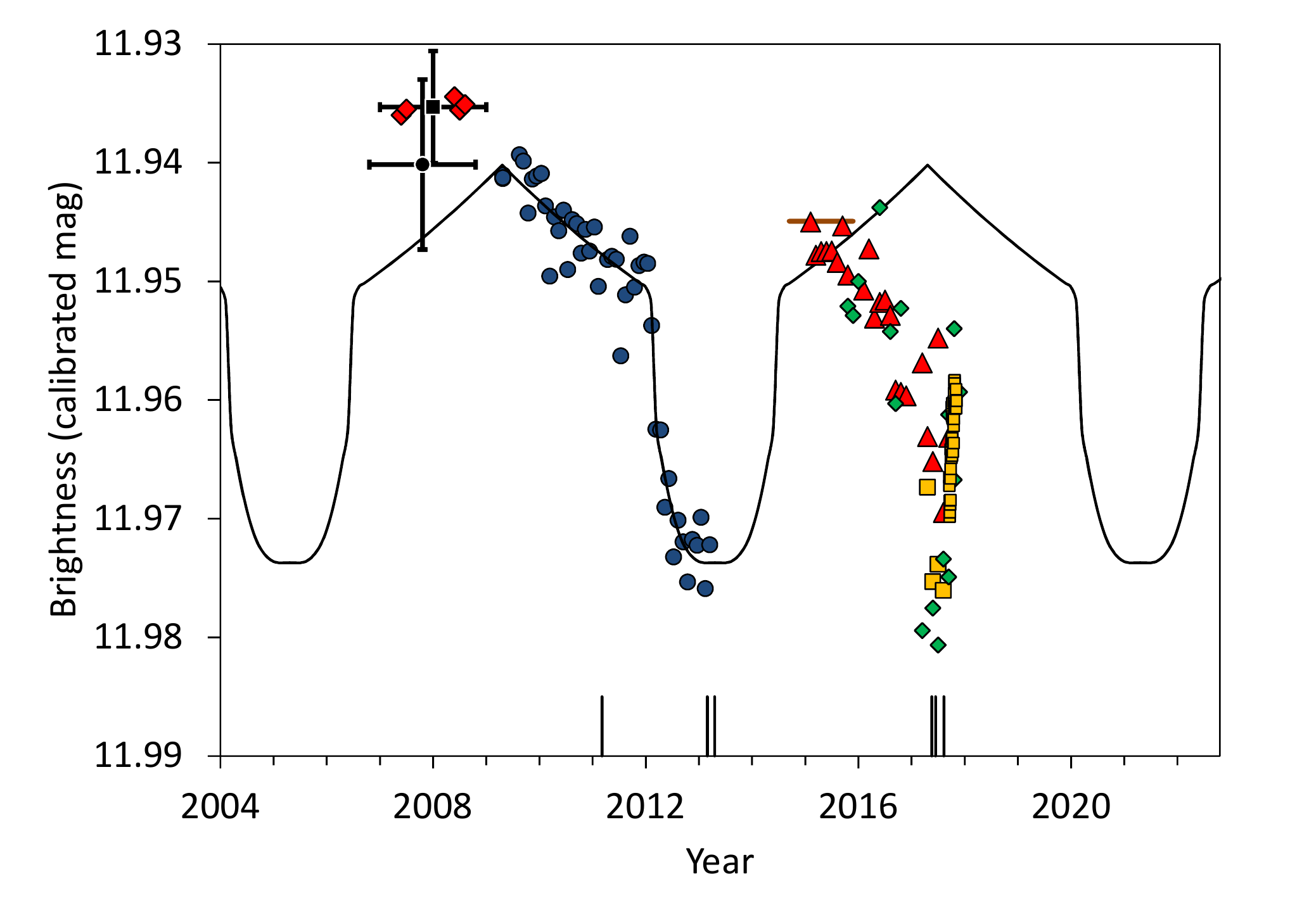}
\includegraphics[width=.5\linewidth]{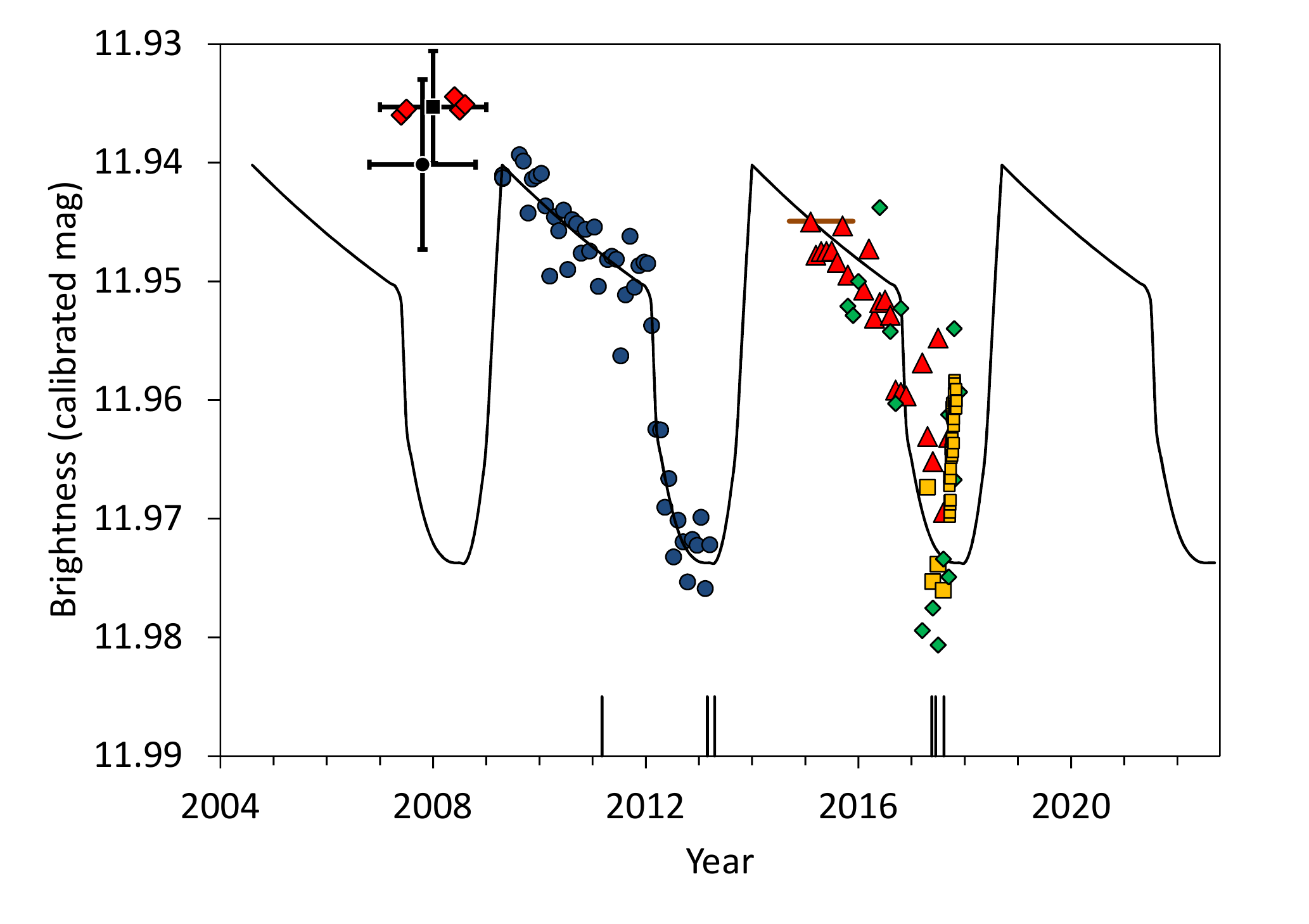}
\caption{\label{fig34}Left: The symmetric model is incompatible with calibrated data. Right: The asymmetric model is incompatible with SuperWASP data (red triangles) which were calibrated to ASAS-V (upper points with uncertainties) and ASAS-I (lower points with uncertainties). The deepest day-long dips are marked with vertical lines.}
\end{figure*}

\subsection{Shortest possible period symmetric model}
Using the highest-quality, uninterrupted FFI data from Kepler (covering $2009-2013$), we create the shortest possible period symmetrical flux model (Figure~\ref{fig2}). The curve has been created by fitting two polynomials to the ``pre-ingress'' and the ``ingress'' time. With the shape of the curve in this model we only aim to determine the duration and amplitude of the signal. This curve covers 8 years in time, compatible to the $\approx8$-year long cycle found by \citet{2017arXiv170807822S}.

For the shortest possible brightness recovery, we stitch a mirror-image of the ingress section immediately to the end of the Kepler data. The shape of this curve is purely phenomenological i.e., it represents the data well, but does not attempt to explain any underlying physical cause. Qualitatively, it resembles a transit-like signal comparable to year-long eclipsing binaries \citep{2016AJ....151..123R}, which could be tested spectroscopically.

The issue with this model is that all datasets other than Kepler do not fit into it (Figure~\ref{fig34}, left). The steady flux from SuperWASP during $2007-2008$ is in disagreement with the expected brightening. Also, the $\approx2$\% dimming seen by ASAS-SN, AAVSO and HAO between 2015 and 2017 is against the expected slight brightening for this time. Clearly, the tested period of 8 years is too long approximately by a factor of two. This can not be rectified by making the period longer or increasing the transit duration.

\subsection{Shortest possible period asymmetric model}
If the model is asymmetric, it can be shorter. We keep the well-defined first half of the model, but replace the ``egress'' with a steep and short curve (Figure~\ref{fig34}, right). Now, the second dimming ($2015-2017$) is well represented in the model. The issue is, however, in the SuperWASP data for the years 2007-2008, which exclude the expected presence of a dip. Instead, there is a hint of a dip at the very beginning of ASAS-V data (Figure~\ref{fig1}) in early 2006, about one year earlier than predicted by the asymmetric model. This could be interpreted as an aperiodic cycle.

We may also challenge the validity of the SuperWASP data, which have been discussed in \citet[][their Figure~7]{2017ApJ...837...85H}. In brief, these data contain 5,355 individual measurements covering 2007.41 to 2007.51 and 2008.36 to 2008.59. The brightness is constant at mmag level during this time. The question of the relative brightness compared to the other datasets remains, as each bandpass is different. SuperWASP ended only 8 months before Kepler started, so that it appears plausible, although uncertain, that they share a similar baseline. If this connection is dropped, we are left with the average flux of ASAS for the respective time. Between 2007.41 and 2008.59, there are 27 individual measurements from ASAS-V with an average, constant brightness of $11.935\pm0.005$\,mag. The result is identical if all 45 measurements for the years 2007 and 2008 are combined. This represents a brightness uncertainty of 0.5\,\% from ASAS-V alone. The dip in the asymmetric model would require a dimming down to at least 11.97 in this bandpass, which is in conflict with the ASAS-V observations by $7\sigma$. When we repeat a similar analysis for the ASAS-I data for $2007-2008$, we obtain an average brightness of $11.94\pm0.007$\,mag, which is consistent with ASAS-V and in conflict with the dip by $4\sigma$.

Therefore, we argue that the asymmetric model can not reproduce the data, because a dip would be expected during 2008 which was not observed. We conclude that neither a symmetric nor an asymmetric model can be periodic on timescales shorter than 10 years (2007-2017). Of course, longer periodic models are still possible.

\begin{figure}
\includegraphics[width=\linewidth]{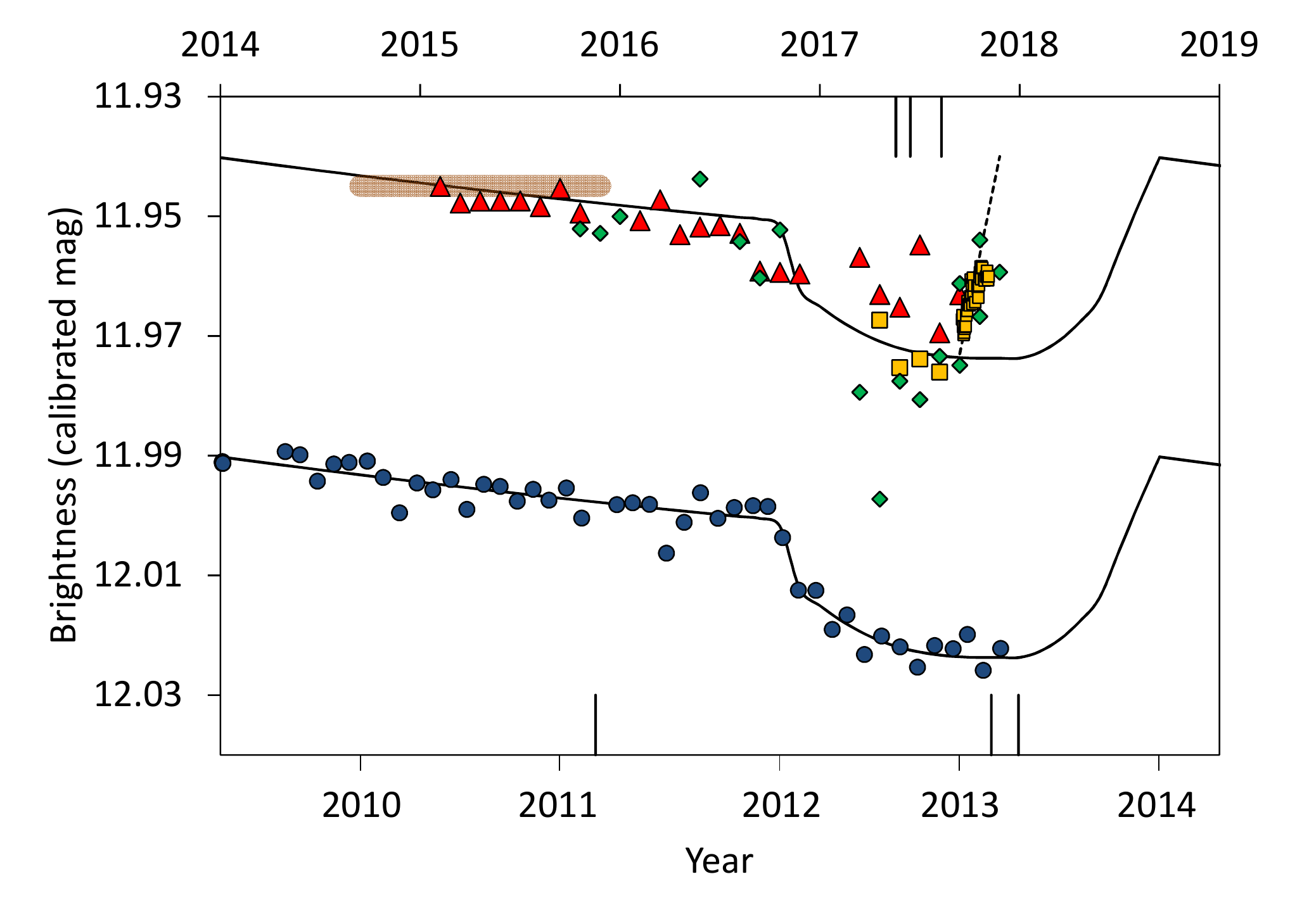}
\caption{\label{fig5}Comparison of Kepler FFI dimming (lower half and ordinate) and recent data (upper part). The cycle shown here is 4.7~years and appears remarkably similar. It is currently (October 2017) unclear if the star is already re-brightening (dashed line).}
\end{figure}

\subsection{Similarity of the dimming slopes 2012 and 2017}
Despite the impossibility of the 4-year symmetric and asymmetric models, the repetition of the dimming slope is remarkable. We achieve the best match to the Kepler FFI data (blue symbols in Figure~\ref{fig5}) with a shift of 4.7~years (1700~days), although shifts as short as 1600~days are visually similar. The latest HAO, AAVSO, ASAS-SN data indicate a steep re-brightening starting in early October 2017. If this trend continues, the star could return back to maximum brightness by the end of 2017 (dashed line in Figure~\ref{fig5}). If the cycle is instead asymmetric, the flux would not follow this line; such a hypothesis can be put to test in late 2017/early 2018.

\citet{2017arXiv171001081S} recently compared the day-long dips from 2013 and 2017 and found similar structures with a period of 1574 days (4.31~years). This is broadly compatible with our time lag. Unfortunately, there is not much data available between late 2016 and early 2017 (because the star was difficult to observe due to it being close to the sun) to track down the exact beginning of the steep decline.

\citet{2017arXiv170807822S} suggest the presence of an 8-year sinusoidal trend based on the ASAS-V and ASAS-I observations (compare light blue symbols in Figure~\ref{fig1}). We can reproduce this cycle with a FFT and a periodogram of the data, but the issue here is that the data covers only 10.93~years, or little more than one such cycle. If an underlying strictly periodic phenomenon exists, and it contains asymmetric structure within, it might be longer and non-sinusoidal. For a robust periodicity analysis, data for several cycles (at least 2-3) are needed

\section{Conclusions}
\citet{2016ApJ...829L...3W} described a number of potential explanations for this object's behavior, such as clouds in the outer Solar System, structure in the interstellar medium (ISM) along the line of sight, natural and artificial material orbiting Boyajian's Star, an intervening object with a large disk or ring, and stellar variations. They conclude that the ISM and intervening disk models are the more plausible ones. Several other groups picked up from there and analyzed individual phenomena in more detail and in the light of new observations. Here we discuss these in the context of our findings.

\citet{2017ApJ...842L...3F} discussed internal stellar effects that potentially explain the flux obstruction. They suggest that magnetic activity, differential rotation, changes in photospheric abundances or just random variation in convective efficiency could produce effects as the ones observed. Such effects, or a combination of them, could be periodic, aperiodic, symmetric or asymmetric. They can however be tested by multicolor photometry, for which observations are ongoing.

As can be seen in Figures~\ref{fig34} and \ref{fig5}, the times of the deepest dips occur just before the subsequent brightening. This might provide interesting evidence that the brightening is caused by internal storage of the blocked energy in the short dips.

\citet{2017MNRAS.471.3680K} posed the question whether the structure in KIC~8462852 could have been caused by matter in the Solar System, considering heliocentric obscuring rings in the outer Solar System that graze the line of sight to the star once per orbit of Kepler. While such a phenomenon is not impossible, it is presently unknown to appear in other stars, and should be periodic at first order. We see the photometry presented in this paper as evidence against such Solar System rings.

\citet{2017arXiv170508427B} proposed that a giant planet orbits KIC~8462852 which hosts a set of rings as well as two massive clouds of trojan asteroids in the planet's Lagrange points on its orbit. Such a complex system could well cause multiple asymmetric dips as observed here, depending on its inclination, eccentricity and overall geometry. Trojans near L4 and L5 can have, in addition, different shapes and sizes, which would be in orbit around their Lagrange points, and thus produce very different transits, or sometimes none at all. Such a system can be constrained with our analysis to a period of at least a decade, and could be probed with transit spectroscopy.

\citet{2017MNRAS.468.4399M} theorized a post-merger return to normal after the ingestion of a planet up to $10,000$ years ago. Gravitational energy released as the body spiraled into the outer layers of the star could have caused a temporary and unobserved brightening, which would explain the (disputed) 100-year dimming. The individual transient dimming events would then be caused by planetary debris from an earlier partial disruption of the same bodies, or due to evaporation and outgassing from a tidally detached moon system. Alternatively, they discuss, similar to the detection paper, that the dimming events could arise from a large number of comet- or planetesimal-mass bodies placed on high-eccentricity orbits by the same mechanism. This model can only weakly be constrained with our analysis, but its effects should fade over time.

Regarding the small-scale ISM structure, \citet{2016ApJ...833...78M} discuss a foreground swarm of comet-like objects or planets crossing the line of sight to the star and its optical companions at approximately 7\,mas per year as a more plausible interpretation than a family of highly eccentric comets orbiting the target star. The swarm may be a free-traveling interstellar group of objects or a belt associated with an additional hypothetical optical source. Again, our data can only weakly constraint this model, but its effects should fade over time. Since the review of \citet{2016ApJ...829L...3W}, the study by \citet{2017ApJ...847..131M} favored circumstellar over ISM obscuration based on new multicolor photometry.

\section{Conclusion}
We have combined and calibrated publicly available data for Boyajian's star. All data are consistent with each other and shows prominent dimming events with an amplitude of 
$\approx4$\,\% in 2012 and 2017, which are remarkably similar in time and depth. Due to the short time in between these events, a symmetric light curve can not be constructed. An asymmetric periodic model is limited by the fact that no dimming was seen during 2007 in ASAS-V ($7\sigma$), ASAS-I ($4\sigma$) and SuperWASP at high confidence. If the dimmings are periodic, their period must exceed ten years. Clearly more observational data are needed to solve this puzzle and we encourage the professional as well as citizen science community to continue their great effort in photometric monitoring.

\acknowledgments
We are thankful to Bruce Gary, Dave Lane, the AAVSO observers and the ASAS team for obtaining and providing data. We thank the referees for their comments which improved the paper.


\begin{thebibliography}{}
\providecommand\natexlab[1]{#1}
\providecommand\JournalTitle[1]{#1}

\bibitem[{{Ballesteros} {et~al.}(2017){Ballesteros}, {Arnalte-Mur},
  {Fernandez-Soto}, \& {Martinez}}]{2017arXiv170508427B}
{Ballesteros}, F.~J., {Arnalte-Mur}, P., {Fernandez-Soto}, A., \& {Martinez},
  V.~J. 2017, \JournalTitle{ArXiv e-prints},
  \href{http://arxiv.org/abs/1705.08427}{{\sffamily arXiv:1705.08427
  [astro-ph.EP]}}

\bibitem[{{Belokurov} {et~al.}(2017){Belokurov}, {Erkal}, {Deason}, {Koposov},
  {De Angeli}, {Evans}, {Fraternali}, \& {Mackey}}]{2017MNRAS.466.4711B}
{Belokurov}, V., {Erkal}, D., {Deason}, A.~J., {et~al.} 2017,
  \href{http://dx.doi.org/10.1093/mnras/stw3357}{\JournalTitle{\mnras}, 466,
  4711}

\bibitem[{{Bodman} {et~al.}(2017){Bodman}, {Quillen}, {Ansdell}, {Hippke},
  {Boyajian}, {Mamajek}, {Blackman}, {Rizzuto}, \&
  {Kastner}}]{2017MNRAS.470..202B}
{Bodman}, E.~H.~L., {Quillen}, A.~C., {Ansdell}, M., {et~al.} 2017,
  \href{http://dx.doi.org/10.1093/mnras/stx1034}{\JournalTitle{\mnras}, 470,
  202}

\bibitem[{{Boyajian} {et~al.}(2016){Boyajian}, {LaCourse}, {Rappaport},
  {Fabrycky}, {Fischer}, {Gandolfi}, {Kennedy}, {Korhonen}, {Liu}, {Moor},
  {Olah}, {Vida}, {Wyatt}, {Best}, {Brewer}, {Ciesla}, {Cs{\'a}k}, {Deeg},
  {Dupuy}, {Handler}, {Heng}, {Howell}, {Ishikawa}, {Kov{\'a}cs}, {Kozakis},
  {Kriskovics}, {Lehtinen}, {Lintott}, {Lynn}, {Nespral}, {Nikbakhsh},
  {Schawinski}, {Schmitt}, {Smith}, {Szabo}, {Szabo}, {Viuho}, {Wang},
  {Weiksnar}, {Bosch}, {Connors}, {Goodman}, {Green}, {Hoekstra}, {Jebson},
  {Jek}, {Omohundro}, {Schwengeler}, \& {Szewczyk}}]{2016MNRAS.457.3988B}
{Boyajian}, T.~S., {LaCourse}, D.~M., {Rappaport}, S.~A., {et~al.} 2016,
  \href{http://dx.doi.org/10.1093/mnras/stw218}{\JournalTitle{\mnras}, 457,
  3988}

\bibitem[{{Boyajian} {et~al.}(2018){Boyajian}, {Alonso}, {Ammerman},
  {Armstrong}, {Asensio Ramos}, {Barkaoui}, {Beatty}, {Benkhaldoun}, {Benni},
  {Bentley}, \& et~al.}]{2018arXiv180100732B}
{Boyajian}, T.~S., {Alonso}, R., {Ammerman}, A., {et~al.} 2018,
  \JournalTitle{ArXiv e-prints},
  \href{http://arxiv.org/abs/1801.00732}{{\sffamily arXiv:1801.00732
  [astro-ph.SR]}}

\bibitem[{{Deeg} {et~al.}(2018){Deeg}, {Alonso}, {Nespral}, \&
  {Boyajian}}]{2018arXiv180100720D}
{Deeg}, H.~J., {Alonso}, R., {Nespral}, D., \& {Boyajian}, T. 2018,
  \JournalTitle{ArXiv e-prints},
  \href{http://arxiv.org/abs/1801.00720}{{\sffamily arXiv:1801.00720
  [astro-ph.SR]}}

\bibitem[{{Foukal}(2017)}]{2017ApJ...842L...3F}
{Foukal}, P. 2017,
  \href{http://dx.doi.org/10.3847/2041-8213/aa740f}{\JournalTitle{\apjl}, 842,
  L3}

\bibitem[{{Gaia Collaboration} {et~al.}(2016){Gaia Collaboration}, {Brown},
  {Vallenari}, {Prusti}, {de Bruijne}, {Mignard}, {Drimmel}, {Babusiaux},
  {Bailer-Jones}, {Bastian}, \& et~al.}]{2016A&A...595A...2G}
{Gaia Collaboration}, {Brown}, A.~G.~A., {Vallenari}, A., {et~al.} 2016,
  \href{http://dx.doi.org/10.1051/0004-6361/201629512}{\JournalTitle{\aap},
  595, A2}

\bibitem[{{Hippke} {et~al.}(2016){Hippke}, {Angerhausen}, {Lund}, {Pepper}, \&
  {Stassun}}]{2016ApJ...825...73H}
{Hippke}, M., {Angerhausen}, D., {Lund}, M.~B., {Pepper}, J., \& {Stassun},
  K.~G. 2016,
  \href{http://dx.doi.org/10.3847/0004-637X/825/1/73}{\JournalTitle{\apj}, 825,
  73}

\bibitem[{{Hippke} {et~al.}(2017){Hippke}, {Kroll}, {Matthai}, {Angerhausen},
  {Tuvikene}, {Stassun}, {Roshchina}, {Vasileva}, {Izmailov}, {Samus},
  {Pastukhova}, {Bryukhanov}, \& {Lund}}]{2017ApJ...837...85H}
{Hippke}, M., {Kroll}, P., {Matthai}, F., {et~al.} 2017,
  \href{http://dx.doi.org/10.3847/1538-4357/aa615d}{\JournalTitle{\apj}, 837,
  85}

\bibitem[{{Katz}(2017)}]{2017MNRAS.471.3680K}
{Katz}, J.~I. 2017,
  \href{http://dx.doi.org/10.1093/mnras/stx1876}{\JournalTitle{\mnras}, 471,
  3680}

\bibitem[{{Lund} {et~al.}(2016){Lund}, {Pepper}, {Stassun}, {Hippke}, \&
  {Angerhausen}}]{2016arXiv160502760L}
{Lund}, M.~B., {Pepper}, J., {Stassun}, K.~G., {Hippke}, M., \& {Angerhausen},
  D. 2016, \JournalTitle{ArXiv e-prints},
  \href{http://arxiv.org/abs/1605.02760}{{\sffamily arXiv:1605.02760
  [astro-ph.SR]}}

\bibitem[{{Makarov} \& {Goldin}(2016)}]{2016ApJ...833...78M}
{Makarov}, V.~V., \& {Goldin}, A. 2016,
  \href{http://dx.doi.org/10.3847/1538-4357/833/1/78}{\JournalTitle{\apj}, 833,
  78}

\bibitem[{Meng {et~al.}(2017)Meng, Rieke, Dubois, Kennedy, Marengo, Siegel, Su,
  Trueba, Wyatt, Boyajian, Lisse, Logie, Rau, \&
  Vanaverbeke}]{2017ApJ...847..131M}
Meng, H. Y.~A., Rieke, G., Dubois, F., {et~al.} 2017,
  \href{http://stacks.iop.org/0004-637X/847/i=2/a=131}{\JournalTitle{The
  Astrophysical Journal}, 847, 131}

\bibitem[{{Metzger} {et~al.}(2017){Metzger}, {Shen}, \&
  {Stone}}]{2017MNRAS.468.4399M}
{Metzger}, B.~D., {Shen}, K.~J., \& {Stone}, N. 2017,
  \href{http://dx.doi.org/10.1093/mnras/stx823}{\JournalTitle{\mnras}, 468,
  4399}

\bibitem[{{Montet}(2017)}]{2017ascl.soft05006M}
{Montet}, B.~T. 2017, {f3: Full Frame Fotometry for Kepler Full Frame Images},
  Astrophysics Source Code Library,
  \href{http://arxiv.org/abs/1705.006}{{\sffamily ascl:1705.006}}

\bibitem[{{Montet} \& {Simon}(2016)}]{2016ApJ...830L..39M}
{Montet}, B.~T., \& {Simon}, J.~D. 2016,
  \href{http://dx.doi.org/10.3847/2041-8205/830/2/L39}{\JournalTitle{\apjl},
  830, L39}

\bibitem[{{Montet} {et~al.}(2017){Montet}, {Tovar}, \&
  {Foreman-Mackey}}]{2017arXiv170507928M}
{Montet}, B.~T., {Tovar}, G., \& {Foreman-Mackey}, D. 2017, \JournalTitle{ArXiv
  e-prints}, \href{http://arxiv.org/abs/1705.07928}{{\sffamily arXiv:1705.07928
  [astro-ph.SR]}}

\bibitem[{{Pojmanski}(1997)}]{1997AcA....47..467P}
{Pojmanski}, G. 1997, \JournalTitle{\actaa}, 47, 467

\bibitem[{{Pojmanski}(2002)}]{2002AcA....52..397P}
---. 2002, \JournalTitle{\actaa}, 52, 397

\bibitem[{{Pollacco} {et~al.}(2006){Pollacco}, {Skillen}, {Collier Cameron},
  {Christian}, {Hellier}, {Irwin}, {Lister}, {Street}, {West}, {Anderson},
  {Clarkson}, {Deeg}, {Enoch}, {Evans}, {Fitzsimmons}, {Haswell}, {Hodgkin},
  {Horne}, {Kane}, {Keenan}, {Maxted}, {Norton}, {Osborne}, {Parley}, {Ryans},
  {Smalley}, {Wheatley}, \& {Wilson}}]{2006PASP..118.1407P}
{Pollacco}, D.~L., {Skillen}, I., {Collier Cameron}, A., {et~al.} 2006,
  \href{http://dx.doi.org/10.1086/508556}{\JournalTitle{\pasp}, 118, 1407}

\bibitem[{{Rodriguez} {et~al.}(2016){Rodriguez}, {Stassun}, {Lund}, {Siverd},
  {Pepper}, {Tang}, {Kafka}, {Gaudi}, {Conroy}, {Beatty}, {Stevens}, {Shappee},
  \& {Kochanek}}]{2016AJ....151..123R}
{Rodriguez}, J.~E., {Stassun}, K.~G., {Lund}, M.~B., {et~al.} 2016,
  \href{http://dx.doi.org/10.3847/0004-6256/151/5/123}{\JournalTitle{\aj}, 151,
  123}

\bibitem[{{Sacco} {et~al.}(2017){Sacco}, {Ngo}, \&
  {Modolo}}]{2017arXiv171001081S}
{Sacco}, G., {Ngo}, L., \& {Modolo}, J. 2017, \JournalTitle{ArXiv e-prints},
  \href{http://arxiv.org/abs/1710.01081}{{\sffamily arXiv:1710.01081
  [astro-ph.SR]}}

\bibitem[{{Schaefer}(2016)}]{2016ApJ...822L..34S}
{Schaefer}, B.~E. 2016,
  \href{http://dx.doi.org/10.3847/2041-8205/822/2/L34}{\JournalTitle{\apjl},
  822, L34}

\bibitem[{{Shappee} {et~al.}(2014){Shappee}, {Prieto}, {Grupe}, {Kochanek},
  {Stanek}, {De Rosa}, {Mathur}, {Zu}, {Peterson}, {Pogge}, {Komossa}, {Im},
  {Jencson}, {Holoien}, {Basu}, {Beacom}, {Szczygie{\l}}, {Brimacombe},
  {Adams}, {Campillay}, {Choi}, {Contreras}, {Dietrich}, {Dubberley},
  {Elphick}, {Foale}, {Giustini}, {Gonzalez}, {Hawkins}, {Howell}, {Hsiao},
  {Koss}, {Leighly}, {Morrell}, {Mudd}, {Mullins}, {Nugent}, {Parrent},
  {Phillips}, {Pojmanski}, {Rosing}, {Ross}, {Sand}, {Terndrup}, {Valenti},
  {Walker}, \& {Yoon}}]{2014ApJ...788...48S}
{Shappee}, B.~J., {Prieto}, J.~L., {Grupe}, D., {et~al.} 2014,
  \href{http://dx.doi.org/10.1088/0004-637X/788/1/48}{\JournalTitle{\apj}, 788,
  48}

\bibitem[{{Simon} {et~al.}(2017){Simon}, {Shappee}, {Pojmanski}, {Montet},
  {Kochanek}, {van Saders}, {Holoien}, \& {Henden}}]{2017arXiv170807822S}
{Simon}, J.~D., {Shappee}, B.~J., {Pojmanski}, G., {et~al.} 2017,
  \JournalTitle{ArXiv e-prints},
  \href{http://arxiv.org/abs/1708.07822}{{\sffamily arXiv:1708.07822
  [astro-ph.SR]}}

\bibitem[{{Wright} \& {Sigurdsson}(2016)}]{2016ApJ...829L...3W}
{Wright}, J.~T., \& {Sigurdsson}, S. 2016,
  \href{http://dx.doi.org/10.3847/2041-8205/829/1/L3}{\JournalTitle{\apjl},
  829, L3}

\end{thebibliography}
\end{document}